\documentclass{aa}

\usepackage{natbib}
\usepackage{amsmath}
\bibpunct{(}{)}{;}{a}{}{,} 
\usepackage{graphics}   
\usepackage{graphicx}   
\usepackage{epstopdf}
\usepackage{longtable}   
\usepackage{url}      
\usepackage{bm}        
\usepackage[varg]{txfonts}
\usepackage{pdflscape}
\usepackage{supertabular}
\usepackage{hyperref}   
\usepackage[svgnames]{xcolor}
\usepackage{longtable}
\usepackage{lscape}
\usepackage{booktabs}
\usepackage{graphicx}
\usepackage{multicol}
\usepackage{placeins}

\usepackage{txfonts}
\usepackage{natbib}
\usepackage{amsmath}
\newcommand\Ms{\ensuremath{M_{\odot}}}
\newcommand\Zs{\ensuremath{\mathrm{Z}_{\odot}}}

\newcommand\Ls{{\ensuremath{\mathrm{L}_{\odot}}}}

\newcommand\Mpy{\Ms\,{\rm yr}{\ensuremath{^{-1}}}}
\newcommand\gva{{\sc Genec}}

\newcommand{\msun}{\ensuremath{M_{\odot}}}

\newcommand{\mdot}{\ensuremath{\dot{M}}}

\begin{document}

   \title{The Evolution of Accreting Population III Stars at 10$^{-6}$ - 10$^3$ \Ms\ yr$^{-1}$}

   \subtitle{}

   \author{Devesh Nandal \inst{1}, Lorenz Zwick\inst{2}, Daniel J. Whalen \inst{3}, Lucio Mayer \inst{4}, Sylvia Ekstr\"om \inst{1}, Georges Meynet
          \inst{1}
          }
\authorrunning{Nandal et al.}
\institute{
Geneva Observatory, Geneva University, CH-1290 Versoix, Switzerland \and Niels Bohr International Academy, Niels Bohr Institute, Blegdamsvej 17, 2100 Copenhagen, Denmark \and Institute of Cosmology and Gravitation, University of Portsmouth, Portsmouth PO1 3FX, UK \and Institut fuer  Astrophysik, Universit\"at Zurich, Winterthurerstrasse 190,  805 Zurich,  Switzerland 
}

   \date{Received; accepted }


  \abstract
   {The first stars formed over five orders of magnitude in mass by accretion in primordial dark matter halos.
}
   {We study the evolution of massive, very massive and supermassive primordial (Pop III) stars over nine orders of magnitude in accretion rate.}   
  {We use the stellar evolution code GENEC to evolve accreting Pop III stars from 10$^{-6}$ - 10$^3$ \Ms\ yr$^{-1}$ and study how these rates determine final masses. The stars are evolved until either the end central Si burning or they encounter the general relativistic instability (GRI).  We also examine how metallicity affects the evolution of the star at one accretion rate.
  }  
   {At rates below $\sim$ $2.5 \times 10^{-5}$ \Ms\ yr$^{-1}$ the final mass of the star falls below that required for pair-instability supernovae.  The minimum rate required to produce black holes with masses above 250 \Ms\ is $\sim 5 \times 10^{-5}$ \Ms\ yr$^{-1}$, well within the range of infall rates found in numerical simulations of halos that cool via H$_2$, $\lesssim 10^{-3}$ \Ms\ yr$^{-1}$.  At rates of $5 \times 10^{-5}$ \Ms\ yr$^{-1}$ to $4 \times 10^{-2}$ \Ms\ yr$^{-1}$, like those expected for halos cooling by both H$_2$ and Ly$\alpha$, the star collapses after Si burning.  At higher accretion rates the GRI triggers the collapse of the star during central H burning.  Stars that grow at above these rates are cool red hypergiants with effective temperatures $log(T_{\text{eff}}) = 3.8$ and luminosities that can reach 10$^{10.5}$ \Ls.  At accretion rates of 100 - 1000 \Ms\ yr$^{-1}$ the gas encounters the general relativistic instability prior to the onset of central hydrogen burning and collapses to a black hole with a mass of $\sim$ 10$^6$ \Ms\ without ever having become a star.  
       }       
   {Our models corroborate previous studies of Pop III stellar evolution with and without hydrodynamics over separate, smaller ranges in accretion rate.  They also reveal for the first time the critical transition rate in accretion above which catastrophic baryon collapse, like that which can occur during galaxy collisions in the high-redshift Universe, produces supermassive black holes via dark collapse.  
      }


   \maketitle
%

\section{Introduction}
   
Primordial (or Pop III) star formation is thought to begin at $z \sim$ 20 - 25 when cosmological  halos reach masses of $\gtrsim$ 10$^5$ \Ms\ and cool by H$_2$ forming in the gas phase (\citealt{stahler1986,bcl99,nu01,abn02} -- see also \citealt{greif14,kg23}).  Although the Pop III initial mass function (IMF) has not been observationally constrained, cosmological simulations suggest that H$_2$ cooling produces primordial stars with masses of a few tens to hundreds of solar masses \citep[e.g.,][]{hos11,hir13,hir15,latif22a}.  These stars are thought to form in small clusters and grow by accretion at rates of up to $10^{-3}$ \Ms\ yr$^{-1}$, which can become clumpy if the accretion disk fragments \citep{clark11,get11,Sakurai2015}.

At slightly later times, the buildup of infrared and Lyman-Werner UV backgrounds by Pop III stars can suppress star formation in other primordial halos through direct photodissociation of H$_2$ and photodetachment of H$^-$, the primary channel through which H$_2$ forms in the gas phase \citep[e.g.,][]{sug14,agar15,latif15a}.  The halos grow to larger masses through mergers and accretion until they reach virial temperatures of 10$^4$ K at masses of 10$^7$ \Ms\ that activate Ly$\alpha$ cooling that triggers catastrophic baryon collapse with infall rates of up to 1 \Ms\ yr$^{-1}$.  In reality, even at these masses and temperatures some H$_2$ survives in the core of the halo by self shielding \citep{prole24a}, so accretion rates at the very center of the halo are mediated by both Ly$\alpha$ and H$_2$ cooling at somewhat lower rates of 0.01 - 1 \Ms\ yr$^{-1}$.  Whether or not this infall eventually produces a single supermassive object or small numbers of massive objects remains unclear because of the short times for which these systems have been evolved to date \citep{rd18b,suaz19,Regan2020}, but recent simulations at lower resolution on timescales of SMS growth and collapse make clear that these halos will host multiple very massive or supermassive stars \citep[SMSs;][]{pat23a}.

Such accretion rates are thought to produce SMSs with masses of a few 10$^4$ - 10$^5$ \Ms\ \citep{Omukai2001,OP2003,Hosokawa2013,umeda2016,woods2017,Lionel2018,herr23a}.  Those that accrete at rates above $\sim$ 0.02 \Ms\ yr$^{-1}$ evolve along cool, red, superluminous Hayashi tracks due to H$^-$ opacity in their outer layers, and those that grow at below this rate migrate to hot, blue compact supergiant tracks \citep{herr23a,Nandal2023a}.  \citet{herr23a} also found this accretion rate to be the lower limit at which the SMS encountered the GRI and collapsed to a BH.  Below it, SMSs collapse because of depletion of core fuel in post main-sequence burning.  It is now thought that collapse can cause some SMSs to explode as highly energetic supernovae (SNe) with energies of up to 10$^{55}$ erg instead of forming BHs \citep{Chen2014,nat22}.  Runaway stellar collisions in dense nuclear clusters at high redshifts can also build up 1000 - 3000 \Ms\ stars \citep{br78,dv09,ksh15,sak17,rein18,boek18}.  Collsions between early galaxies can even gravitationally torque gas at solar metallicities into supermassive nuclear disks (SNDs) that may be prone to radial GR instabilities and collapse at rates of up to 10$^5$ \Ms\ yr$^{-1}$ \citep{may10,Mayer2019,Zwick2023}, calling into question if stars can even form at such accretion rates \citep{Lionel2020,Lionel2021b}.  SMSs are currently the leading contenders for the origin of the first quasars at $z >$ 6 \citep[e.g.,][]{dm12,smidt18,latif22b} and less massive BHs found at $z =$ 10.1 and 10.6 in the {\em James Webb Space Telescope} ({\em JWST}) JADES and CEERS surveys \citep[UHZ1 and GN-z11;][]{akos23,maio23}.

No single study has ever considered the entire range of accretion rates relevant to Pop III star evolution, from those mediated by H$_2$ cooling at low temperatures \citep{Murphy2021} to those associated with the merger scenario of SMBH formation, in which the GRI might envelop the protostar prior to main sequence burning.  Here, we extend previous studies to model Pop III star evolution over nine decades in accretion rate, from 10$^{-6}$ - 10$^3$ \Ms\ yr$^{-1}$, in 1 dex steps with the Geneva Stellar Evolution Code (\gva).  For clarity, throughout this paper, we refer to protostars as accreting objects powered primarily by contraction, and stars as those mainly driven by nuclear reactions. In Section~\ref{2:models}, we describe our \gva\ models.  We present  Hertzsprung-Russell diagrams for our protostars and stars in Section~\ref{3: evolution} and their properties as a function of accretion rate in Section~\ref{4: properties}. The final masses of the accreting objects are discussed in Section~\ref{5: masses} and the impact of changing metallicity is covered in Section~\ref{6: metal}. Future studies are considered in Section~\ref{7: conc}.

\section{Models}\label{2:models}

We assume that accretion proceeds at constant rates.  Other studies have begun to examine SMS evolution in the cosmological flows that create them (\citealt{tyr21a,tyr24a} -- see also \citealt{Sakurai2015,sakurai2016a}).  However, they generally find outcomes similar to those for stars that evolve at the equivalent average rate and it would not have been possible to systematically investigate the large range in accretion rate we consider here with such models.  We also assume that accretion proceeds at the same rate regardless of if the star evolves onto a hot, blue track and becomes a significant source of ionizing UV flux that could photoheat infalling gas and reduce its flow rate.  This is not a serious limitation, as recent cosmological simulations of Direct Collapse Black Hole (DCBH) formation that include radiation from the progenitor find that its UV flux at most reduces accretion rates by 50\% so they can grow at the rates we consider here \citep[e.g.,][]{latif21a}.  We also assume 'cold' accretion in which the accretion luminosity is assumed to be radiated away, with no transport of entropy into the interior of the star.  \citet{Hosokawa2013} considered both extremes, in which the accretion luminosity was also taken up into the star, and found only minor differences in evolution over early ranges in mass that disappeared above a few thousand solar masses.

Our stars are evolved from the pre-main sequence (pre-MS) phase to the end of core silicon burning (with the exception of 10$^{-2}$ M$_{\odot}$yr$^{-1}$ model which is evolved until the later stage of core helium burning, Y$_{c} = 0.4$) with \gva\ \citep{Eggenberger2008}.  They have initial homogeneous chemical compositions of $X =$ 0.7516, $Y =$ 0.2484, and metallicities $Z =$ 0, consistent with \citet{Murphy2021}.  The 0.1 \Ms\ yr$^{-1}$ models at $Z =$10$^{-6}$ and \Zs\ have the same chemical composition used in \citet{Nandal2023b}. All models begin their evolution as fully convective hydrostatic seeds with different initial masses; models with accretion rates from 10$^{-6}$ \Ms\ yr$^{-1}$ to 0.1 \Ms\ yr$^{-1}$ start as 2 \Ms seeds and models with accretion rates beyond 0.1 \Ms\ yr$^{-1}$ start as 10 \Ms\ seeds. The choice of different initial seed masses is to facilitate numerical converge during the early preMS phase (first 50 years) and such a choice does not affect the evolution and final fates of models. We also include deuterium with a mass fraction $X_2$ = $5 \times 10^{-5}$, as in \citet{Bernasconi1996,Behrend2001,Lionel2018}.  The equations of state, nuclear reaction rates, opacities and the same those used in \citet{Ekstrom2012} and \citet{Nandal2023b}. Numerical convergence poses significant challenges in models at high accretion rates. Particularly, models with accretion rates of \(10^{2}\) and \(10^{3} \, M_{\odot}\) per year encounter considerable difficulties. To secure the most accurate physical solutions, we systematically decrease the time steps between fully converged solutions, spanning several hours to days. Further reduction in time steps would necessitate a level of numerical precision that surpasses the capabilities of current computational machinery. Moreover, the models with the highest accretion rates incorporate approximately 400 shells, a number chosen to balance the trade-off between numerical stability and the fidelity of the physical representation. Despite these efforts, the spatial resolution of these models remains a potential area for enhancement and such models may exhibit sudden change in radius due to spurious numerical defects.The onset of GRI was determined using the criteria described in Saio et al (in prep) which corresponds to the linear adiabatic pulsational analysis based on the work by \citet{Chandrasekhar1964}. The instability is determined when the frequency of adiabatic pulsations is negative and the amplitude of pulsations grows in the interior of the star. The equations 2-6 in \citet{Nagele2022} describe the criteria governing the instability in detail. Once the accreting models are found to reach the GRI according to the criteria described in \citet{Nagele2022}, the models are post processed in the linear adiabatic pulsational code (Saio et al (in prep)) to determine the final mass. When a convective core is present, we use a step overshoot $\alpha_{ov} = 0.1$. We do not include rotation in our models because they merit separate studies, and we also ignore mass loss due to line-driven winds.  Previous studies have shown that mass loss is negligible at metallicities below 0.01 \Zs\ because H and He lines do not present enough opacity to outgoing photons \citep{kud00,Vink01,bhw01}. They would be present in our solar metallicity models but would be overcome by accretion at the rates we consider.

\section{HR Diagrams}\label{3: evolution}

\begin{figure*}
\includegraphics[width=20.0cm]{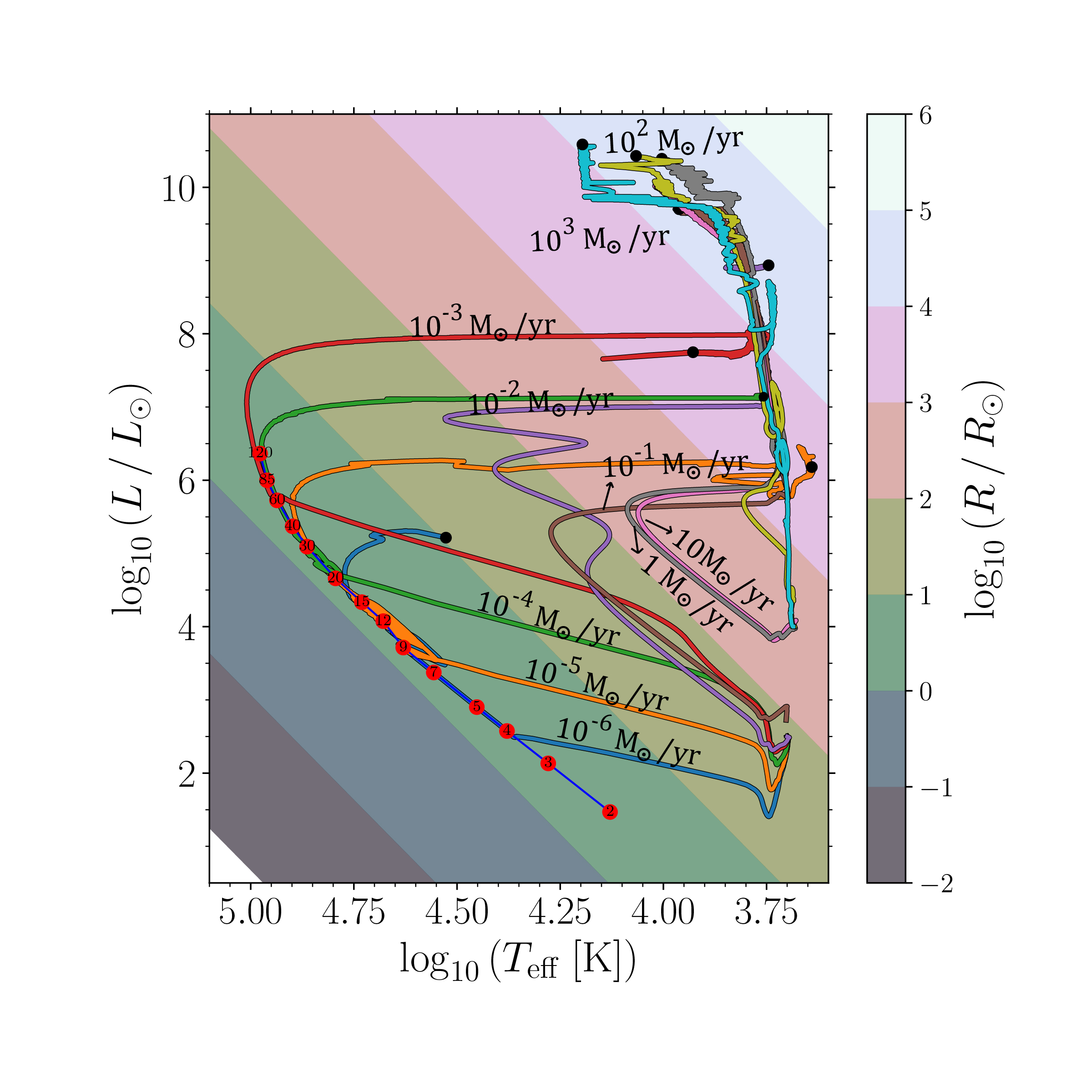}
\caption{HR diagram of massive and supermassive stars at zero metallicity at accretion rates of $10^{-6}$ \Ms\ yr$^{-1}$ to $10^{3}$ \Ms\ yr$^{-1}$. The background colors indicate radii in units of solar radius.  The blue line represents the $Z =$ 0 ZAMS track with red dots marking masses in solar units. The tracks are labelled by accretion rate in black. The black dots indicate the end point of computation which corresponds to end of core Silicon burning for accretion rates up to $10^{-2}$ \Ms\ yr$^{-1}$ and the onset of GR instability for higher accretion rates.   
}
\label{BIG}
\end{figure*} 

\begin{figure*}
\includegraphics[width=18.4cm]{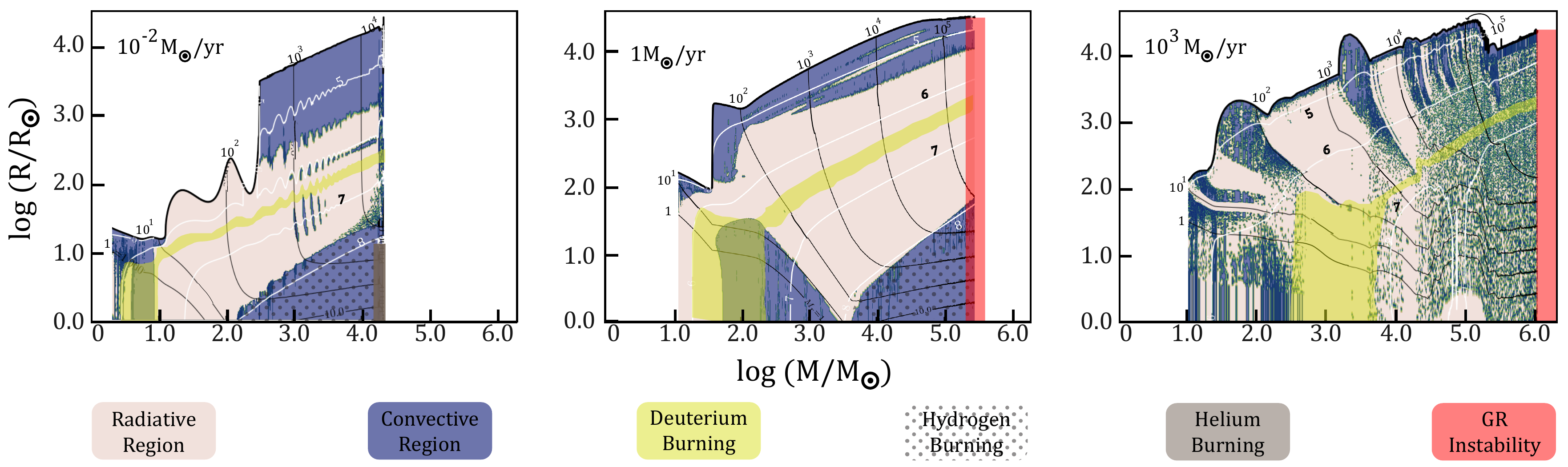}
\caption{Evolution of the total radius as a function of growing mass at accretion rates of 10$^{-2}$ \Ms\ yr$^{-1}$ (left panel), 1 \Ms\ yr$^{-1}$ (centre panel) and 10$^3$ \Ms\ yr$^{-1}$ (right panel) shown in  Kippenhahn diagrams.}
\label{Kipp}
\end{figure*}

\begin{figure*}   
\centering
\includegraphics[width=9cm]{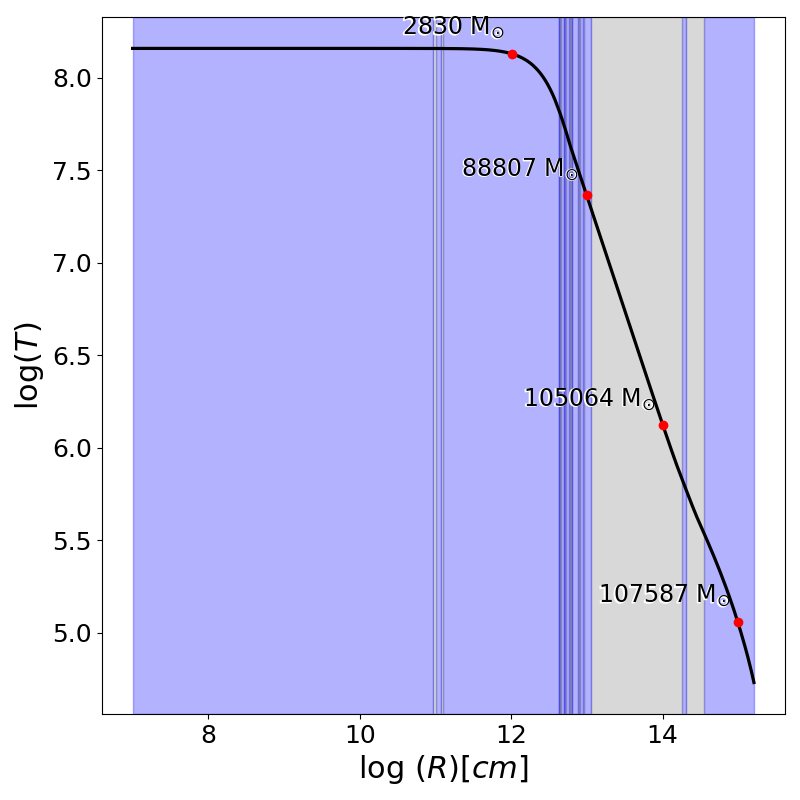}    
\includegraphics[width=9cm]{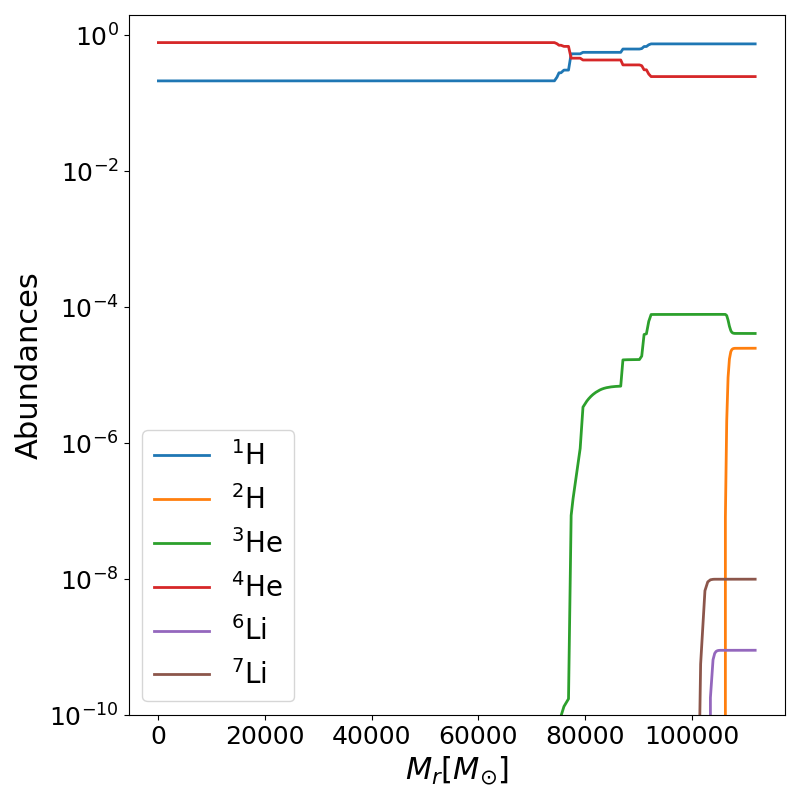}
\caption{{\it Left panel:}Log T versus log R plot for the 0.1 \Ms/yr model at the end of its evolution upon encountering GR instability at a mass of 108000 \Ms. The blue zones indicate convective zones and grey zones correspond to radiative zones. The enclosed mass is indicated at different radii (see the numbers at red points). Most of the mass (105000 \Ms) is within a radius of 10$^{13}$ cm or 14.0 $R_\odot$ that represents 1\% of the total radius. Only a fraction of mass (still 23000 \Ms) in the outer envelope implying the star is extremely bloated.
      {\it Right panel:} The plot depicts the change in abundance of various elements inside the 0.1 \Ms/yr model upon encountering the GR instability when central mass fraction of hydrogen is 0.244.}
\label{TR}
\end{figure*}    

\begin{figure*}   
\centering
\includegraphics[width=18.4cm]{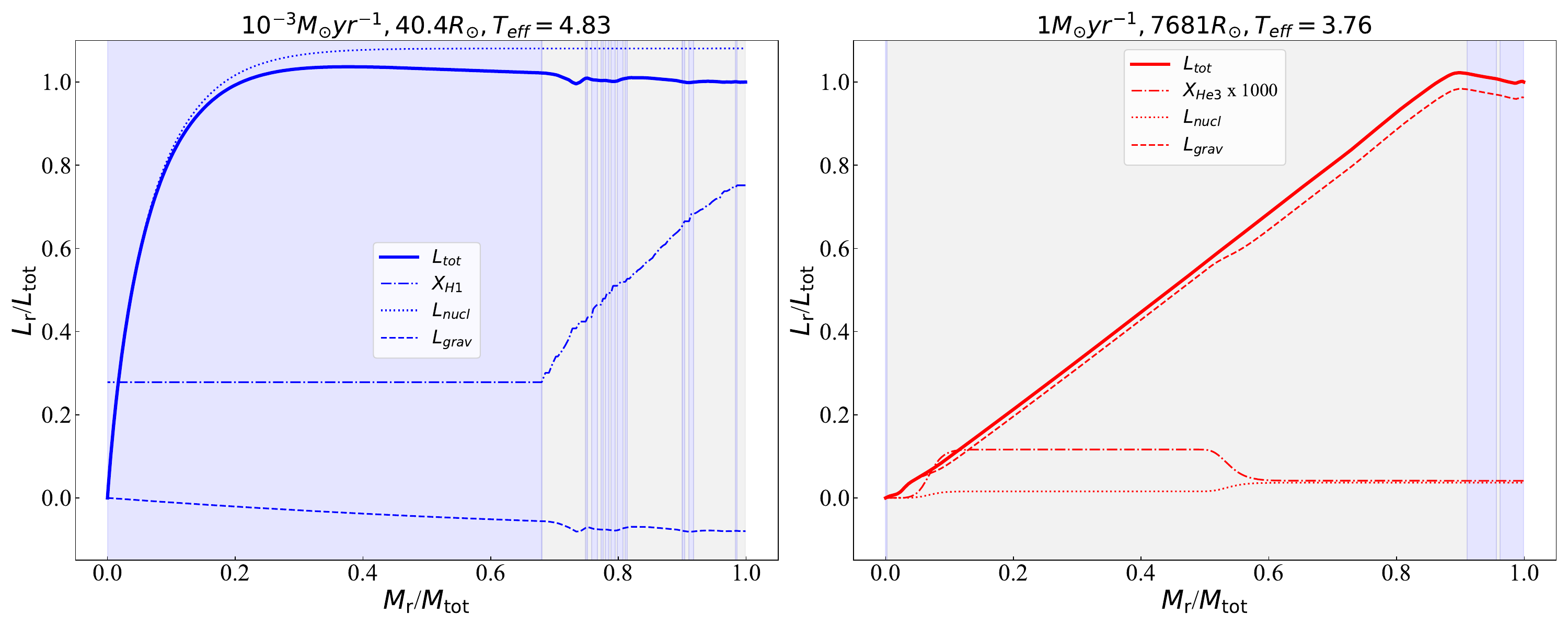}    
\caption{Variation of Luminosity (in mass fraction) as a function of mass coordinate inside two accreting models compared at same total luminosity. {\it Left:}  total (solid), nuclear (dotted) and gravitational (dashed) luminosities versus enclosed mass fraction for the 10$^{-3}$ \Ms\ yr$^{-1}$ star at a mass of 2072 \Ms\ and total luminosity of $6.9 \times 10^7$ \Ls. {\it Right:} Same as on the left but for a pre-MS 1.0 \Ms\ yr$^{-1}$ star at a mass of 2337 \Ms\ and same total luminosity (in this case powered only by gravitational contraction). The dot-dashed lines show mass fractions for H (left) and $^{3}$He (right) in the star. The shaded blue and grey zones depict the convective and radiative regions respectively.
}
\label{BR}
\end{figure*}   

\begin{figure*}
\includegraphics[width=18.4cm]{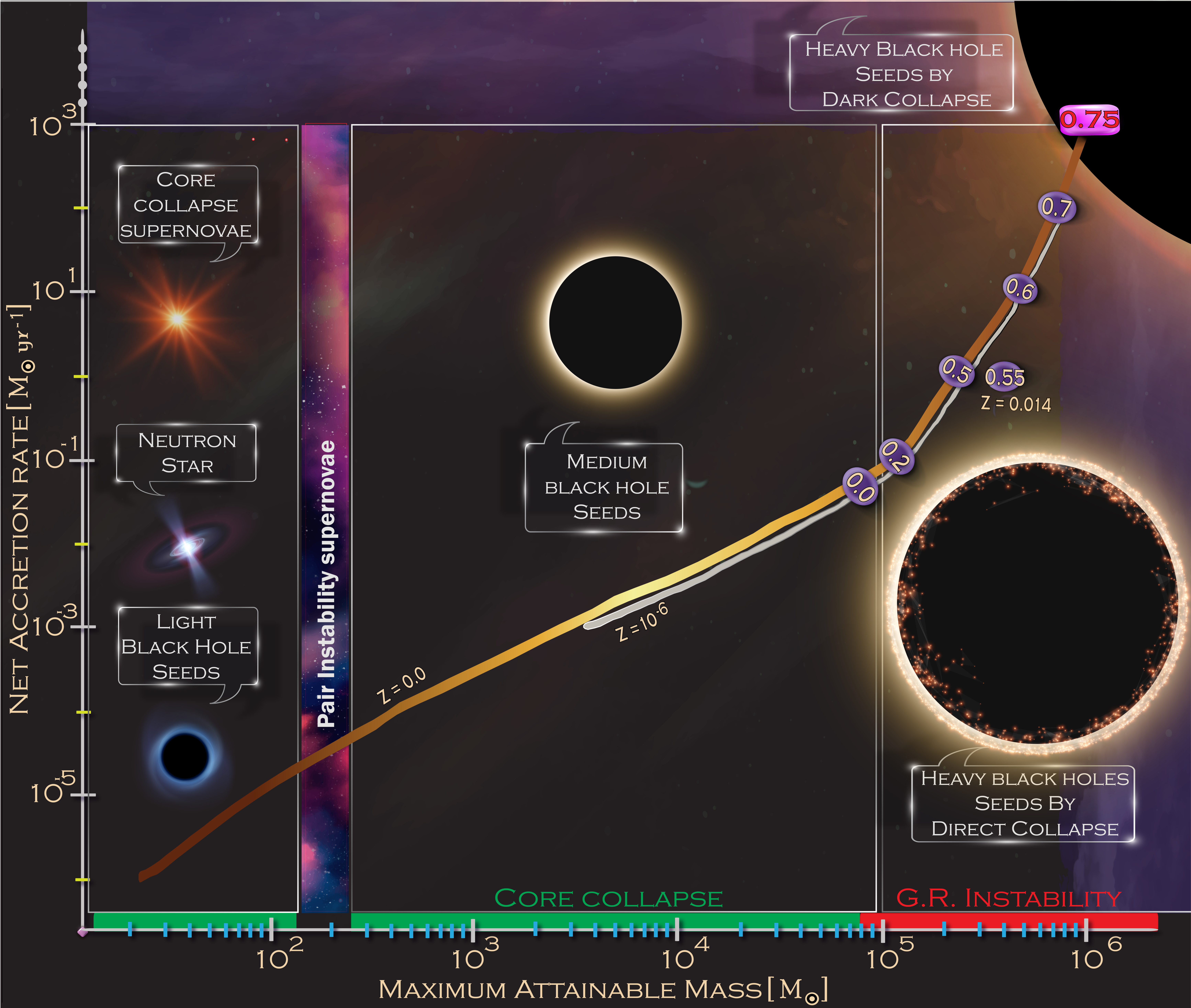}
\caption{Final mass versus accretion rate for $10^{-6}$ \Ms\ yr$^{-1}$ to $10^{3}$ \Ms\ yr$^{-1}$. The gold line is for zero metallicity models, the silver line is for $Z = 10^{-6}$ \Zs\ and the single dot at 1 \Ms\ yr$^{-1}$ is the solar metallicity model. The green bar marks the final mass range for which collapse occurs after core Si burning and the red bar is the mass range for which the GRI triggers collapse during central H burning.  The vertical pair instability strip spans 150 - 250 \Ms\ and is for the non accreting models while 7 - 150 \Ms\ progenitors produce neutron stars or stellar-mass BHs. The oval shapes with numbers correspond to the the central mass fraction of hydrogen at the time GRI is triggered.}
\label{BIG2}
\end{figure*}    

We show evolution tracks for our stars from 10$^{-6}$ - 10$^3$ \Ms\ yr$^{-1}$ in 1 dex increments in the HR diagram in Figure~\ref{BIG}.  They fall into two basic categories, hot compact blue stars at accretion rates below $\sim$ 0.02 \Ms\ yr$^{-1}$ and cool red hypergiants on the Hayashi track above these rates \citep{Nandal2023a,herr23a}.  Below 10$^{-2}$ \Ms\ yr$^{-1}$, we observe several features.  First, as previously found in models at solar metallicity, the higher the accretion rate the larger the mass \(M_{\rm J}\) at which the track joins the Zero Age Main Sequence (ZAMS).  For instance, Figure~\ref{BIG} shows that this mass is about 3.5 \Ms\ at 10$^{-6}$ \Ms\ yr$^{-1}$ and 60 \Ms\ at 10$^{-3}$ \Ms\ yr$^{-1}$.  Second, core H burning mostly occurs in the blue region of the HR diagram. However, as with classical tracks of core H-burning models, the birthlines curve rightward. This phase extends over a broader range of effective temperatures as the accretion rate increases.  Third, in our models the most luminous blue stars reach maximum luminosities of about 10$^8$ \Ls.

A rough estimate of $M_{\rm J}$ can be obtained by equating formation time (t$_{form} =$ M$^*$/$\dot{M}_{\rm acc}$) with Kelvin-Helmholtz (KH) contraction time. The KH time can be found from the radius and luminosity of the star at the joining point so
\begin{equation}
 \log \left( \frac{M_{\rm J}}{{\rm M}_\odot} \right) = -0.1726 + \frac{3}{2}\log \left( \frac{L}{{\rm L}_\odot} \right) - 2 \log T_{\rm eff} - \log \dot{M}_{\rm acc}.
\label{eq:MJ}
\end{equation}
Here, the accretion rate is in \Ms\ yr$^{-1}$ and the effective temperature is in Kelvin. Table~\ref{tab:KH} lists the input values used to estimate $M_{\rm J}$ and the values derived from the numerical simulations. It is observed that for 10$^{-3}$ \Ms\ yr$^{-1}$ there is a significant discrepancy between the analytic and numerical results\footnote{The actual age when $M_{\rm J}$ is 60 \Ms\ is greater by nearly a factor 2 than the estimate from its KH time.}.  Note that Equation~\ref{eq:MJ} assumes that the pressure is dominated by an ideal gas when in reality radiation also contributes at higher masses. When radiation pressure is significant, a larger fraction of the energy released during contraction is used to increase the internal energy in order to maintain hydrostatic equilibrium. This gives a shorter Kelvin-Helmholtz timescale at a given mass. This implies that to accommodate a given formation timescale, a larger mass has to be considered. This explains the discrepancy in \(M_{\rm J}\) between Equation~\ref{eq:MJ} and our \gva\ models.


The evolution tracks of stars evolving along the Hayashi line before the onset of the core H burning transition from blue to red at luminosities that decrease as the accretion rate increases from 10$^{-2}$ - 10$^{-1}$ \Ms\ yr$^{-1}$. At higher accretion rates, the luminosity at the crossing towards the Hayashi line stabilizes at 10$^{5.7}$ - 10$^6$ \Ls.  The tracks subsequently ascend the Hayashi line, reaching luminosities of up to 10$^{10}$ \Ls.  They evolve to bluer colors towards the end of core H burning. Finally, the stars can attain much larger luminosities, up to 10$^{10.5}$ \Ls\ at log($T$) $=$ 4.00 - 4.25. 

\section{Pop III SMS Properties}\label{4: properties}

\subsection{Internal Structures}   

We show Kippenhan diagrams of the structural evolution of the 10$^{-2}$, 1 and 1000 \Ms\ yr$^{-1}$ models in Figure~\ref{Kipp}.  Several interesting features emerge:

\begin{enumerate}

\item For all three models there is a point where the radius increases almost vertically on the diagram, indicating rapid expansion in short times. Rapid expansion prevents significant mass increase over those times and occurs at log($M/\Ms)$) $\sim$ 2.5, 1.5, and 1.5 at the three rates, respectively.
    
\item An outer convective zone develops, which diminishes in spatial extent as the accretion rate increases.

\item The isomass lines reveal that the majority of the star's mass undergoes contraction.

\item At 1 \Ms\ yr$^{-1}$, as seen in the middle panel, there are two distinct phases with a convective core. The first phase coincides with core deuterium burning and the second phase is driven by hydrogen burning.
    
\item At 1000 \Ms\ yr$^{-1}$ the core is nearly fully convective after the first 100 years of evolution. Conditions are never met for core hydrogen burning.\end{enumerate}

\subsection{Chemical Structures}

We show temperatures and elemental abundances for the 0.1 \Ms\ yr$^{-1}$ star at the end of its evolution in Figure~\ref{TR}, when it encounters the GRI at a central H mass fraction of 0.2. It has a surface region where the chemical composition is still primordial.  The temperature at the bottom of the outer convective envelope is low enough to prevent the destruction of fragile species such as $^{2}$H, $^{3}$He and Li isotopes. At this point the star has a luminosity of 10$^{9.6}$ \Ls.  In principle, if our models are accurate and surface abundances can be determined spectroscopically, some of these SMSs might retain surface abundances of light isotopes identical to those in primordial material.  If the star loses its outer envelope at the time of collapse, this mass would have little if any $^{12}$C, $^{14}$N or $^{16}$O, just $^1$H and $^4$He.  In contrast, \citet{Nandal2023b} found that Pop III stars accreting at much lower rates, $10^{-4} - 10^{-3}$ \Ms\ yr$^{-1}$, that reach final masses of 1000 - 3000 \Ms, do not encounter the GRI but collapse after Si burning.  As they transition to core He burning they become nearly fully convective, which leads to strong internal mixing and a surface enriched first by $^4$He and then $^{14}$N, $^{16}$O and $^{12}$C. 

\begin{table}
    \caption{$M_{\rm J}$ from Equation~\ref{eq:MJ} and \gva.  Masses are in the units of solar masses.}    
    \centering
    \begin{tabular}{cccccc}
    \hline
     $\dot{m}_{\rm acc}$   &  $\log L$/\Ls\ & $\log T_{\rm eff}$ & $M_{\rm J}$ (Eq. 1)  & $M_{\rm J}$ (\gva) \\ 
     \hline
     10$^{-6}$ & 2.30  & 4.325 & 4.24   &  3.5    \\
     10$^{-5}$ & 3.60  & 4.610 & 10.00 &  8.0    \\ 
     10$^{-4}$ & 4.75  & 4.800 & 22.00 &  20.0  \\     
     10$^{-3}$ & 5.75  & 4.950 & 35.68 &  60.0  \\      
     \hline
    \end{tabular}
    \label{tab:KH}
\end{table}

\subsection{Maximum Luminosities}

Figure~\ref{BR} shows that the maximum luminosities reached by accreting blue stars is about 10$^8$ \Ls\ at effective temperatures near 60000 K (\(\lg T_{\rm eff} \sim 4.75\)). Red stars can exceed 10$^{10.5}$ \Ls\ at effective temperatures below 18000 K (\(\lg T_{\rm eff} < 4.25\)).  As shown in the left panel of Figure~\ref{BR}, stars at the extremes of blue and red in the HR diagram have significantly different structures. Consider the 10$^{-3}$ \Ms\ yr$^{-1}$ star at a mass of 2072 \Ms\ at \(\lg T_{\rm eff} = 4.83\), which is undergoing core H burning.  Its central H mass fraction is 0.30 with the convective core covering about 70\% of the total mass.  The 1 \Ms\ yr$^{-1}$ star at the same luminosity and similar mass, 2337 \Ms, is in the HR diagram's redder region and has not started hydrogen burning, so its luminosity is primarily powered by contraction. Unlike the 10$^{-3}$ \Ms\ yr$^{-1}$ star, the ratio \(L_r/L_{\rm tot}\) in the 1 \Ms\ yr$^{-1}$ star is nearly linear with enclosed mass fraction (right panel of Figure~\ref{BR}). Over enclosed mass fractions of 0.1 - 0.5, the isotope \(^{3}\)He is formed but becomes depleted in the core. This star has an outer convective envelope that is about 10\% of its total mass, most of the star is radiative as shown in the center panel of Figure~\ref{Kipp}.

\section{Final Stellar Masses}\label{5: masses}

\begin{table}
\centering
\caption{Final properties of the stars in this study. $X_{\rm c, final}$ is the core H mass fraction at the end of the life of the star. M$_{core}$ represents the mass of the convective core.}
\label{tab:1}

\resizebox{\linewidth}{!}{

\begin{tabular}{ *9l }    \toprule
$\dot{M} $     &    End-point     &      $t_{\rm H}$      &   $t_{red}/t_{tot}$       &      $M_{\rm final}$     &      $X_{\rm c, final}$     &     $M_{\rm core}$      \\
\Mpy      &                &      Myr                  &    	                        &      \msun     &         &     \msun      \\\midrule
0.000001      &     End Si-burn           &      22.5                  &    	0.002                        &      22.5     &    0     &     5.2      \\\midrule
0.00001      &     End Si-burn           &      8.40                  &    	0.002                        &      86     &    0     &     13      \\\midrule
0.0001      &     End Si-burn           &      4.36                  &    	0.002                        &      436     &    0     &     121      \\\midrule
0.001      &     End Si-burn           &      2.796                  &    	0.003                        &      3053     &    0     &     660      \\\midrule
0.010        &     Mid He-burn (0.49Y$_{c}$)           &      2.261                  &    	0.998                        &      22600     &    0     &     9912      \\\midrule
0.050        &     GR instab. (0.99Y$_{c}$)           &      1.680                  &    	0.903                        &      80206     &    0     &     44792      \\\midrule
0.075        &     GR instab.           &      1.345                  &    	0.826                        &      100972     &    0.16     &     62017      \\\midrule
0.1          &     GR instab.          &      1.040                 &    	0.630                        &      108000     &    0.244     &     67330      \\\midrule
1             &     GR instab.          &      0.214                  &    	0.230                        &      214100     &    0.525     &     60090      \\\midrule
10           &     GR instab.          &      0.045                  &    	0.438                        &      449700     &    0.619     &     47560      \\\midrule
100           &     GR instab.          &      0.007                  &    	0.194                        &      697200     &    0.721     &     7406      \\\midrule
1000           &     GR instab.          &      0.001                  &    	0.118                        &      1027570     &    0.751     &     0      \\\midrule
\bottomrule

\end{tabular}
}
\label{tab:fin}
\end{table}

Final masses and fates for the stars are shown versus accretion rate in Figure~\ref{BIG2} \citep[compare to Figure~2 of][]{hw02}. Although we do not follow collapse in \gva, it is caused by the depletion of core fuel and formation of an iron core, the onset of pair-creation by thermal photons in the core at the expense of pressure support there, or the GRI, in which thermal photons contribute to gravity in the core via the stress-energy tensor and trigger violent pulsations that produce collapse.  We list which process ends the life of the stars, and at what stage of evolution, in Table~\ref{tab:fin}.

\subsection{Nuclear Limit}

The nuclear mass limit is reached when the formation timescale of the accreting star becomes equal to its total nuclear lifetime, when post-MS burning creates an iron core.  At our lowest accretion rate the star reaches a mass of about 30 \Ms, well above that required for an iron core to form, so its final mass is determined by the nuclear limit.  At higher accretion rates the star can reach greater masses before encountering the nuclear limit but at some point it will encounter an instability that causes it to collapse before an iron core can form.  At these rates other limits impose a final mass on the star as explained below.

\subsection{Pair Instability Limit}

Massive stars encounter the pair instability \citep[PI;][]{rs67,brk67} when temperatures in their cores exceed 10$^9$ K and thermal photons have enough energy to freeze out as electron - positron pairs, usually during central O or Si burning.  The dip in pressure support due to the loss of these photons causes the core to contract, which can trigger explosive thermonuclear burning of O and Si.  For non-accreting 140 - 260 \Ms\ Pop III stars the energy released can completely unbind the star, with no compact remnant left behind \citep{hw02,wet12b,wet12a}.  At 100 - 140 \Ms\ the PI produces consecutive pulses that eject large masses from the star but do not fully disrupt it, producing bright transients and leaving behind an iron core that later collapses to a BH \citep[the pulsational pair-instability, or PPI;][]{wet13d,chen14a,wbh07,w17,lnb19}.  Rotation can build up more massive He cores at lower total masses, enabling such stars to encounter the PI at 85 \Ms\ \citep{cw12,smidt14a,cw14a}.  At masses above 260 \Ms, non-accreting Pop III stars encounter the photodisintegration instability, in which thermal photons in the core preferentially crack He nuclei apart instead of creating positron-electron pairs \citep{hw02}.  This process results in the prompt collapse of the core to a BH.  

While 85 - 260 \Ms\ stars can be built up by accretion rates of $2 \times 10^{-5}$ \Ms\ yr$^{-1}$ - $8 \times 10^{-5}$ \Ms\ yr$^{-1}$ in our models, it is not clear how accretion would affect the onset of the PI, and we cannot answer this question here because \gva\ lacks the ability to switch to hydrodynamics to capture the explosion.  Likewise, we cannot properly model the PPI because \gva\ cannot reproduce the mass ejections that would disrupt accretion onto the star or therefore evolve the star in a self-consistent manner.  Nevertheless, the mass limits for the PI and PPI for non-accreting Pop III stars are useful and reasonable approximations to the PI mass limit for accreting stars.

\subsection{GRI Limit}

\begin{table}
\centering
\caption{Comparison of final masses at the onset of the GRI in units of 10$^5$ \Ms: \citealt{umeda2016} (UM16), \citealt{woods2017} (W17), \citealt{Lionel2018} (HLE18) and \citealt{Lionel2021a} (HLE21).}
\label{tab:3}
\begin{tabular}{rccccc}    
\toprule
\mdot\ \hspace{0.2in}  &   UM16   &   W17    &   HLE18   &   HLE21   &   this work  \\
\midrule
0.1 \Ms\ yr$^{-1}$        &    1.20    &    1.65    &     0.75     &        -        &       1.08      \\
1    \Ms\ yr$^{-1}$        &    3.50    &    2.60    &     1.16     &     2.29     &       2.04      \\
10  \Ms\ yr$^{-1}$        &    8.00    &    3.20    &     2.61     &     4.37     &       4.49      \\
\bottomrule  
\end{tabular}
\end{table}

\begin{figure*}
\includegraphics[width=9.5cm]{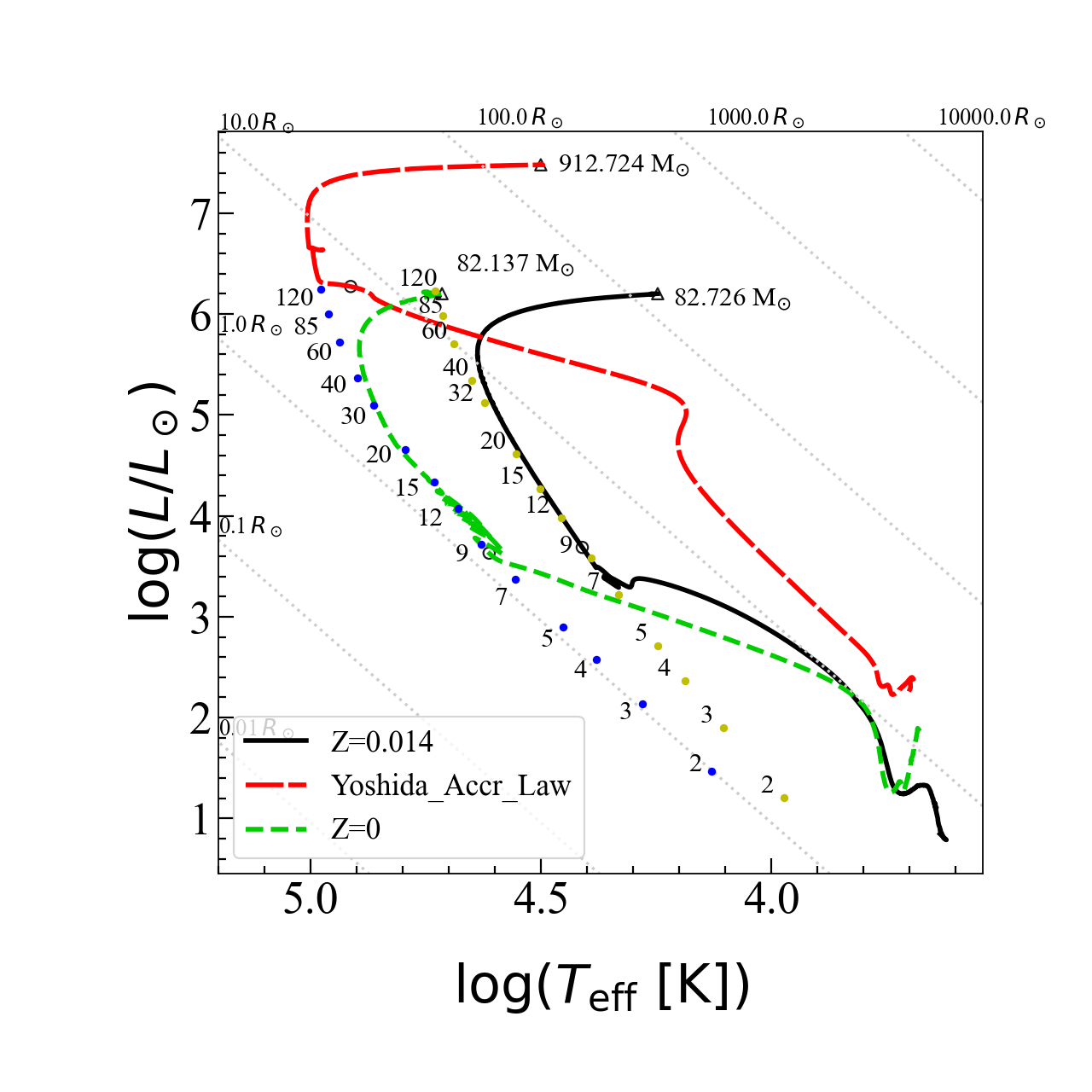}    
\includegraphics[width=9.5cm]{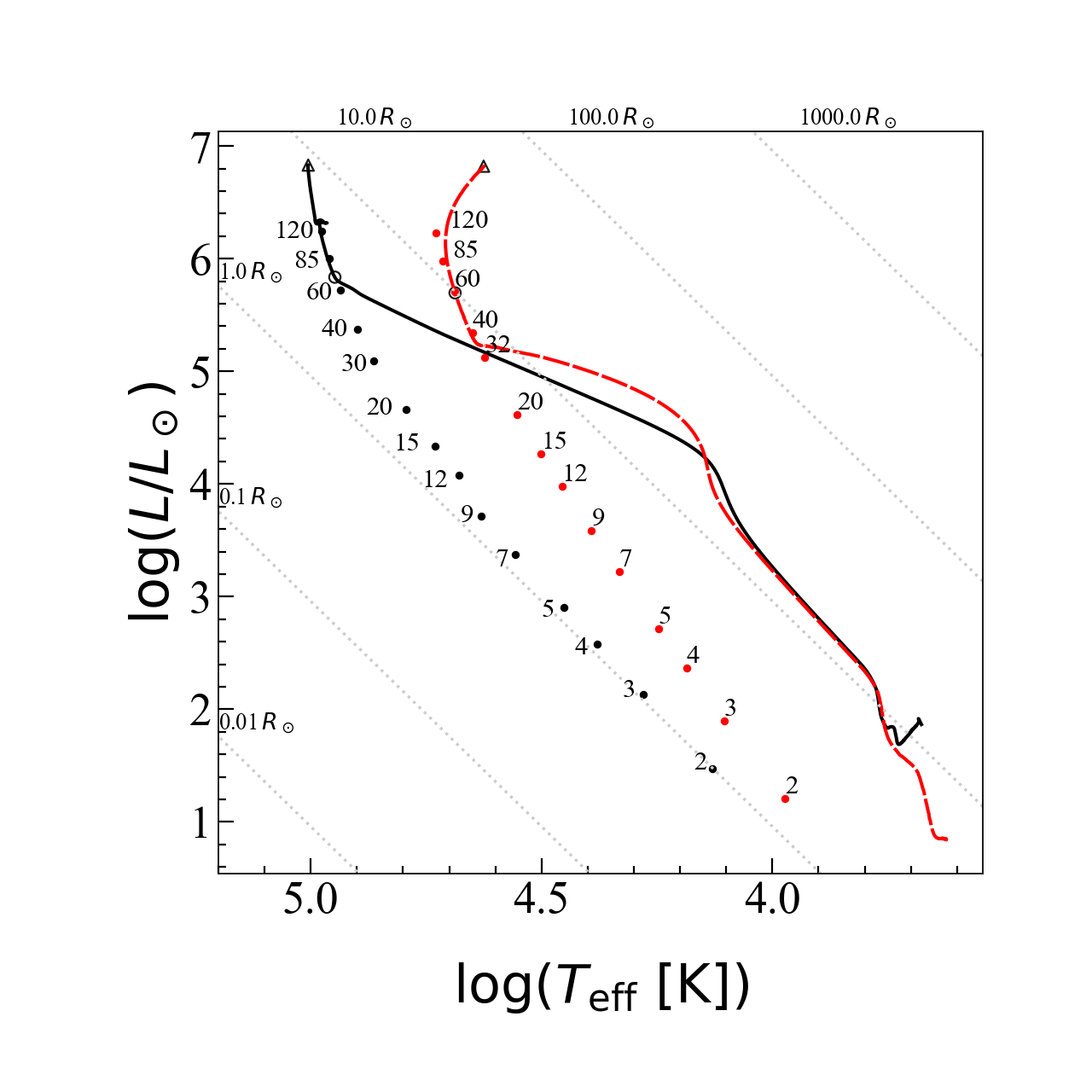}
\caption{Birthlines for a few accretion rates and metallicities. {\em Left:} the dashed red line is for the variable accretion rate from a hydrodynamical simulation of a Pop III star by \citet{Yoshida2006}. The solid black track is the 10$^{-5}$ \Ms\ yr$^{-1}$ solar metallicity model. The short dashed green track is the 10$^{-5}$ \Ms\ yr$^{-1}$ zero-metallicity model.  The black circles and triangles mark the beginning and end of core hydrogen burning, respectively.  Masses at the end of hydrogen burning are shown at the end of each track.  The solid blue and yellow numbered dots and their masses (in \Ms) are non-accreting ZAMS stars at zero and solar metallicity from \citet{Murphy2021} and \citet{Ekstrom2012}, respectively.  The grey dotted lines are iso-radii for the indicated radii.  {\em Right:}  birthlines for Pop III (black) and solar metallicity (red) stars with a constant accretion rate of 10$^{-3}$ \Ms\ yr$^{-1}$.  The solid black and red numbered dots and their masses (in \Ms) are again the stars at zero and solar metallicity from \citet{Murphy2021} and \citet{Ekstrom2012}, respectively, for comparison.
}
\label{COMP0_S}
\end{figure*} 

\begin{figure}
\includegraphics[width=9.0cm]{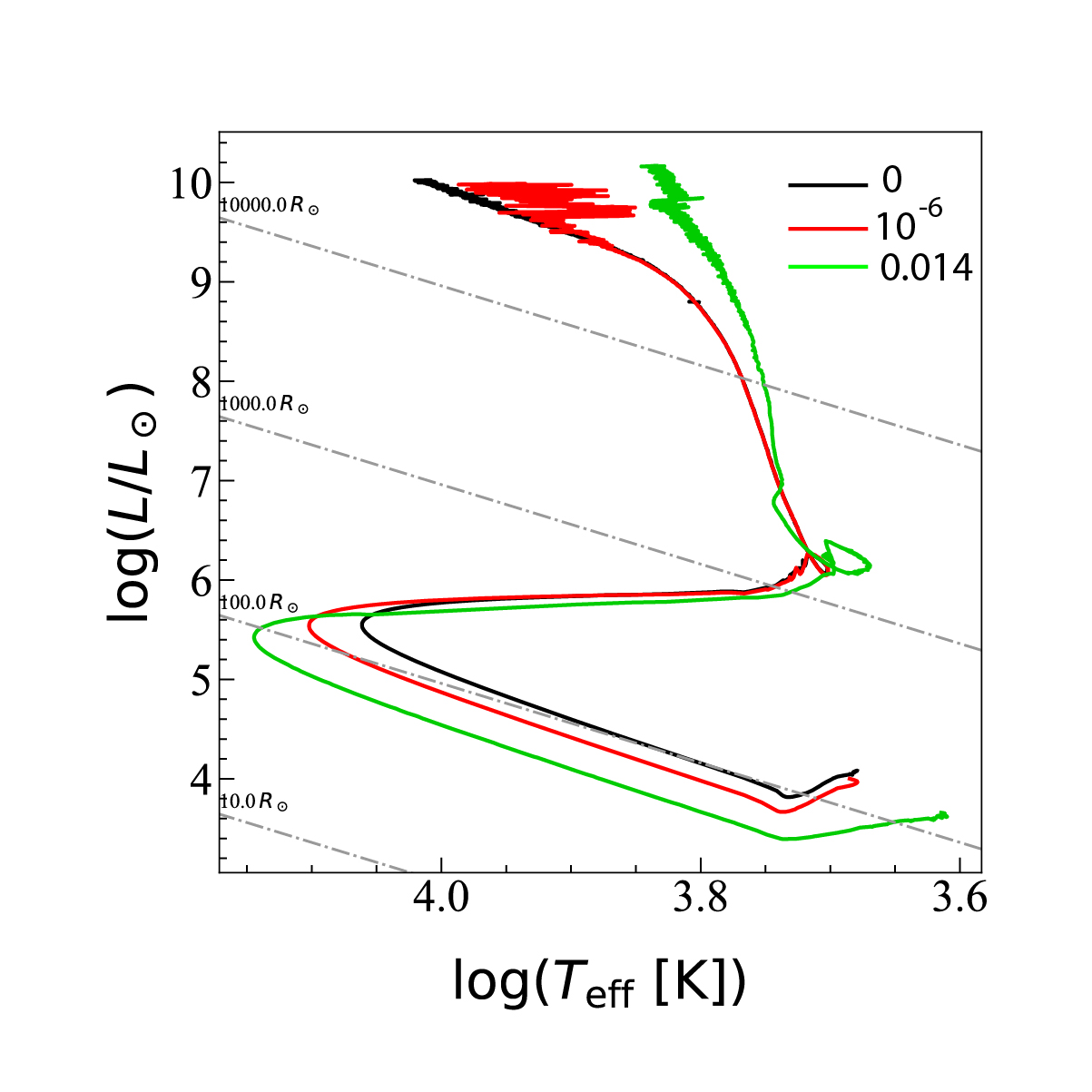}
\caption{HR diagram of 1 \Ms\ yr$^{-1}$ SMS models at three metallicities of Z = 0 (black line), Z = 10$^{-6}$ (red) and Z = solar (green).}
\label{Zeffect2}
\end{figure}   

\begin{figure*}
\includegraphics[width=19.0cm]{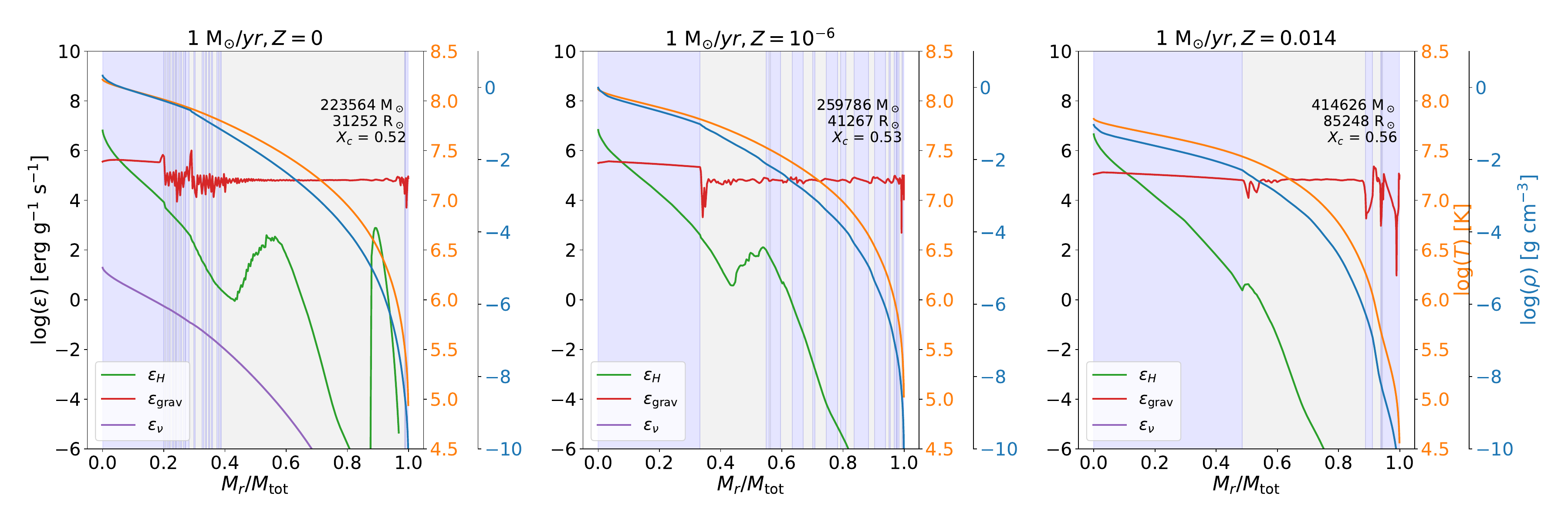}
\caption{Densities, temperatures and energy generation rates as functions of enclosed mass and radius for 1 \Ms\ yr$^{-1}$ stars at zero (left), $10^{-6}$ (centre) and solar (right) metallicities. The light blue and light grey regions indicate the convective and radiative zones respectively.}
\label{Zeffect1}
\end{figure*}

At accretion rates above a few 10$^{-2}$ \Ms\ yr$^{-1}$ the star encounters the GRI, typically during central H burning.  We find the minimum accretion rate required to trigger the GRI to be  0.05 \Ms\ yr$^{-1}$, when the star is at the onset of core He burning at a mass of 80206 \Ms. Higher rates produce the GRI at earlier times in the MS.  In Figure~\ref{BIG2}, we show the central hydrogen mass fraction at points where GRI occurs.  At accretion rates beyond 100 \Ms\ yr$^{-1}$, the GRI happens before core H burning begins, leading to the birth of a dark collapse black hole \citep{Mayer2019}.

We list final SMS masses due to collapse via the GRI in previous studies in Table~\ref{tab:3} for several accretion rates.  They can vary for several reasons.  First, codes can linearize the Tolman-Oppenheimer-Volkoff (TOV) equation in a number of ways that can cause differences in mass at which the GRI triggers pulsations in the stars (in codes with hydrodynamics) or at which the stars satisfy semianalytical criteria for collapse (in codes without hydrodynamics).  Second, criteria for collapse by the GRI in the absence of hydrodynamics vary by code.  Finally, violent pulsations triggered by the GRI in codes with hydrodynamics usually cause the collapse of the star at lower masses than those at which criteria for collapse are met in codes without hydrodynamics.  For example, we obtain a higher minimum accretion rate at which SMSs encounter the GRI than \citet{herr23a}, who found it to be $\sim$ 0.02 \Ms\ yr$^{-1}$ because their code \citep[MESA;][]{paxt11,paxt13,paxt18} transitions to hydrodynamics at the onset of pulsations.  Nevertheless, as shown in Table~\ref{tab:3}, our final masses are generally within 50\% agreement with those in previous studies and differ by at most less than a factor of two.

\subsection{Analytical Fits to Final Masses}

The final stellar masses in Figure~\ref{BIG2} can be fit by linear regressions in log accretion rate, $\dot{M}$, applied separately to the two ranges in rate separated by 0.05 \Ms\ yr$^{-1}$, above which the star collapses via the GRI and below which the star runs up against the nuclear or PI mass limits:
\begin{equation}
    M_{\text{max}} = 
    \begin{cases} 
        10^{5.832} \times \dot{M}^{0.765}, & \text{for } \dot{M} \leq 5.00 \times 10^{-2} \\
        10^{5.301} \times \dot{M}^{0.261}, & \text{for } \dot{M} > 5.00 \times 10^{-2}.
    \end{cases}
    \label{eq:Max_formula}
\end{equation}
Here, $\dot{M}$ is in \Ms\ yr$^{-1}$. This relationship, which is based on our \gva\ stellar evolution models, provides an approximation of a star's maximum mass and is consistent with the nuclear, PI, and GR mass limits. 

\section{Metallicity}\label{6: metal}

\subsection{Low Accretion Rates}

The left panel of Figure~\ref{COMP0_S} shows birthlines for accreting stars at solar and primordial compositions.  The primordial birthline is based on accretion rates from a hydrodynamical simulation by \citet{Yoshida2006}. For solar metallicity, we used a constant accretion rate of 10$^{-5}$ \Ms\ yr$^{-1}$.  Comparison of these two birthlines offers insights into stellar evolution in primordial haloes versus the solar neighborhood. During the initial evolution phase before contracting to the ZAMS, the track for the accreting Pop III protostar shifts to lower effective temperatures than those with solar metallicity. This shift is primarily due to the accretion rate rather than metallicity. This is evident in the right panel of Figure~\ref{COMP0_S}, where birthlines for Pop III and solar metallicity stars with constant accretion rates of 10$^{-3}$ \Ms\ yr$^{-1}$ are shown. Both tracks align up to a luminosity of 4.3 in units of log (L/\Ls), suggesting that metallicity has minimal impact. In the left panel of Figure~\ref{COMP0_S}, the birthline with solar metallicity and a lower accretion rate reaches the ZAMS at about 7 \Ms. In contrast, the Pop III star at the higher accretion rate reaches the ZAMS at a much higher mass, around 120 \Ms.  This difference is due to accretion rate, a phenomenon previously discussed by \citet{Lionel2016}, because a higher accretion rates lead to larger masses within a KH time.

The right panel of Figure~\ref{COMP0_S} shows that the mass at which the birthline reaches the ZAMS is influenced by metallicity.  For a given accretion rate, $M_{\rm J} \sim$ 32 \Ms\ at solar metallicity compared to 60 \Ms\ at \(Z=0\). This difference is due to \(Z=0\) stars being more compact because of lower opacities and the absence of CNO elements, which are crucial for catalyzing hydrogen burning in the core. Their smaller radii at a given mass and luminosity lead to longer KH timescales and thus more time to accumulate mass.  The birthlines for the \(Z=0\) and solar metallicity stars at the same accretion rate in the left panel of Figure~\ref{COMP0_S} are quite similar, suggesting that accretion rate plays a more significant role than initial metallicity. Notably, the tracks overlap in the less luminous and redder regions of the HR diagram. Differences begin to appear as the star contracts towards blue regions. Furthermore, both birthlines reach the ZAMS at almost the same mass, with the Pop III star being slightly more massive. Their masses at the end of the MS are also similar. The mass at the end of the MS varies significantly with accretion rate. At lower accretion rates, the final mass is around 83 \Ms\ while it exceeds 900 \Ms\ at higher ones.

\subsection{High Accretion Rates}

Figure~\ref{Zeffect2} shows evolutionary tracks for stars at various metallicities at an accretion rate of 1 \Ms\ yr$^{-1}$.  Overall, the differences in tracks are relatively minor.  At luminosities above 10$^6$ \Ls, during core H burning, the tracks shift toward the red at higher metallicities.  This shift occurs because at higher metallicities the internal opacity of the star is larger and expands the star's outer layers, causing cooler temperatures at optical depths of about 2/3.  This effect changes the star's radius and effective temperature but not its luminosity.  At luminosities below 10$^6$ \Ls, the tracks shift towards the blue with increasing metallicity. This shift arises because at initial stages of accretion the higher opacities affects more of the star's total mass, not just its outer layers, which in turn reduces the luminosity for a given mass. Therefore, as metallicity increases, the tracks are not only shifted to the right but also towards lower luminosities. 

Figure~\ref{Zeffect1} shows some properties of the same stars at the end of their evolution. Comparison of the plots shows several interesting points:

\begin{enumerate}

\item The energy released per unit time and mass due to contraction is nearly constant through the interior at all metallicities, averaging around \(10^5\) ergs g\(^{-1}\) s\(^{-1}\).
    
\item The final mass is higher for greater metallicities. Consequently, the lifetimes vary with metallicity, being almost twice as long at solar metallicity as at zero metallicity.
    
\item Stars have larger radii at higher metallicities. The solar metallicity SMS has more than twice the radis of the Pop III SMS.
    
\item There is a notable bump in nuclear energy production rate at a mass fraction of 0.6 that is more pronounced in the Pop III SMS than in the solar metallicity star. 
    
\item Central temperatures are higher in the Pop III model, which is a direct result of the absence of CNO elements.

\end{enumerate}
Even during core H burning, these stars primarily generate energy through contraction. The longer lifetime of the solar metallicity SMS stems from its larger convective core, which prolongs core H burning and provides more time for accretion. The larger radius is a result of higher opacities in the outer layers.

\section{Conclusion}\label{7: conc}

We have explored the evolution of Pop III stars accreting at rates of 10$^{-6}$ - 10$^3$ \Ms\ yr$^{-1}$.  We find that 
\begin{enumerate}

\item Stellar evolutionary tracks can broadly categorized by accretion rate, with a critical rate of approximately 0.025 \Ms\ yr$^{-1}$  below which stars migrate towards blue tracks and above which follow the Hayashi line prior to core H burning.  This result is consistent with \citet{herr23a} and \citet{Nandal2023b}.
    
\item Blue stars reach bolometric luminosities up to 10$^8$ \Ls\ at high effective temperatures and red stars reach 10$^{10.5}$ at lower temperatures.
    
\item Below 10$^{-2}$ \Ms\ yr$^{-1}$, the mass at which tracks join the ZAMS increases with accretion rate. Core H burning occurs in the blue region of the HR diagram.
    
\item Above 0.05 \Ms\ yr$^{-1}$, corresponding to a final mass of 84600 \Ms, stars collapse via the GRI during core H burning.  At extreme accretion rates of 100 - 1000 \Ms\ yr$^{-1}$ the gas encounters the GRI prior to the onset of H burning and collapses to a BH without forming a star (dark collapse).  The GRI thus imposes an upper limit on mass on stars of approximately 10$^6$ \Ms.
    
\item Metallicity significantly influences the birthlines of stars. The evolutionary tracks shift towards the blue at lower luminosities and towards the red at higher luminosities with increasing metallicity. For a constant accretion rate of 1 \Ms\ yr$^{-1}$, higher metallicity results in a larger total mass at the end of evolution, increased stellar radii, and enhanced nuclear energy production.
    
\end{enumerate}
We found that Pop III stars accreting even at modest rates of 10$^{-3}$ - 10$^{-4}$ \Ms\ yr$^{-1}$, like those found in numerical simulations of primordial halos cooled by H$_2$, can reach masses of 300 - 3000 \Ms.  In reality, ionizing UV feedback from the star would almost certainly terminate its growth at well below these masses by photoevaporating its accretion disk \citep[and even driving all the gas from the halo;][]{wan04,ket04}.  Nevertheless, stars could still reach such masses in the multiphase environments of more massive halos that cool by both H$_2$ and Ly$\alpha$, in which radiation from the star can reduce but not fully suppress accretion.

As noted earlier, because \gva\ does not capture pulsations triggered by the GRI, our final SMS masses should be taken to be upper limits because previous studies have shown that such pulsations can lead to the collapse of the core at lower masses than those found here.  This raises the greater point that pulsations due to other mechanisms could arise well before those due to the GRI but have not been found by studies to date because they arise on timescales that are much shorter than the nuclear burning timescales on which the stars are advanced.  In principle, such pulsations could lead to other instabilities that collapse or explode the star at lower masses than those found here.  They would also clearly improve their prospects for detection at high redshifts by periodically brightening the star and allowing it to be found in searches for transients \citep{nag23a}.  But they could also end the growth of the star, especially if mass ejections become comparable to lower accretion rates.  We will investigate the existence of such pulsations in future work.

Our models exclude rotation and magnetic fields, which have been shown to be important to the evolution of massive stars.  Although SMSs are expected to be slow rotators due to the $\Omega-\Gamma$ limit \citep{mm00}, \citet{hle18a} found the effect of rotation on the lifetimes of such objects to merit further study. The presence of internal magnetic fields, especially during the early formation phase, has been shown to promote the growth of metal-enriched SMSs \citep{hir23}. The role of the Taylor-Spruit dynamo \citep{egg22} in the transport of angular momentum and chemical species inside accreting SMSs has not yet been investigated.  

At accretion rates of 100 - 1000 \Ms\ yr$^{-1}$ the equations of stellar structure solved by \gva\ and other codes such as MESA may no longer properly describe the buildup of the object, and a general relativistic hydrodynamics code coupled to nuclear burning may be required to determine the final fate of the gas.  Stellar evolution codes produce a series of quasi-static snapshots of the structures of stars when in reality the growing object is the product of converging flows that compress gas on much smaller time steps.  However, to evolve these accreting flows in \gva\ we had to artificially restrict the time steps to hours or minutes to obtain numerical convergence, and evolving the equations of stellar structure on these timescales is not radically different than a full fluid flow solution.  Our constraints on accretion rates for dark collapse, while approximate, are reasonable.  

The \citet{Mayer2019} merger scenario for the prompt formation of 10$^9$ \Ms\ SMBHs at $z >$ 6 invokes collisions between galaxies at $z \sim$ 8 - 10 that are already at solar metallicities to justify central infall rates of 10$^2$ - 10$^5$ \Ms\ yr$^{-1}$, but we have only considered primordial compositions here.  Higher metallicities could change the critical accretion rate for dark collapse or even trigger explosions that are driven by rapid proton captures or explosive CNO burning \citep{nag23b}.  However, we found that the object encounters the GRI before central H burning in the absence of C, N, and O. At higher metallicities 
CNO burning occurs at lower temperatures and the situation may be different.  

\begin{acknowledgements}
The authors thank Dr Lionel Haemmerl\'e for his help and interesting discussions.
D.N., S.E. and G.M. have received funding from the European Research Council (ERC) under the European Union's Horizon 2020 research and innovation programme (grant agreement No 833925, project STAREX). L.Z. acknowledges support from ERC Starting Grant No. 121817-BlackHoleMergs. G.M. has received funding from the SNF project 200020-212124. 

\end{acknowledgements}

\bibliographystyle{aa}
\bibliography{ms}

\end{document}